\documentstyle[11pt,aaspp4]{article}

\def\prop{\propto}
\newbox\grsign \setbox\grsign=\hbox{$>$} \newdimen\grdimen \grdimen=\ht\grsign
\newbox\simlessbox \newbox\simgreatbox \newbox\simpropbox
\setbox\simgreatbox=\hbox{\raise.5ex\hbox{$>$}\llap
     {\lower.5ex\hbox{$\sim$}}}\ht1=\grdimen\dp1=0pt
\setbox\simlessbox=\hbox{\raise.5ex\hbox{$<$}\llap
     {\lower.5ex\hbox{$\sim$}}}\ht2=\grdimen\dp2=0pt
\setbox\simpropbox=\hbox{\raise.5ex\hbox{$\prop$}\llap
     {\lower.5ex\hbox{$\sim$}}}\ht2=\grdimen\dp2=0pt

\def\simless{\mathrel{\copy\simlessbox}}

\def\bin{\,{\rm bin}}

\def\cminvsq{\,{\rm cm}^{-2}}

\def\s{\,{\rm s}}
\def\ms{\,{\rm ms}}

\def\ph{\,{\rm ph}}
\def\secinv{\,{\rm s}^{-1}}

\def\Fp{F_{\rm p}}
\def\tr{t_{\rm r}}
\def\td{t_{\rm d}}

\newcommand{\bez}{\begin{eqnarray*}}
\newcommand{\eez}{\end{eqnarray*}}
\newcommand{\be}{\begin{equation}}
\newcommand{\ee}{\end{equation}}
\newcommand{\beq}{\begin{eqnarray}}
\newcommand{\eeq}{\end{eqnarray}}
\newcommand{\bc}{\begin{center}}
\newcommand{\ec}{\end{center}}

\lefthead{Stern, Poutanen, and Svensson}
\righthead{Brightness dependent properties of  gamma-ray bursts}

\begin{document}

\title{Brightness Dependent Properties of Gamma Ray Bursts}

\author{Boris Stern \altaffilmark{1,2}, Juri Poutanen\altaffilmark{2,3}, 
and Roland Svensson\altaffilmark{2}
\altaffiltext{1}{Institute for Nuclear Research, Russian Academy of Sciences
Moscow 117312, Russia, stern@al20.inr.troitsk.ru}
\altaffiltext{2}{Stockholm Observatory, S-133 36 Saltsj\"obaden, Sweden,
stern@astro.su.se, juri@astro.su.se, svensson@astro.su.se}
\altaffiltext{3}{Uppsala Observatory, Box 515, S-75120 Uppsala, Sweden} 
} 


\begin{abstract}

Brightness dependent correlations have been
conclusively detected in the average time profiles of gamma-ray bursts
(GRBs).
Determining the time constants of the stretched exponential  
slopes of the average peak-aligned time profile of GRBs 
as a function of the  peak brightness, 
we find that the post-peak slope shows time 
dilation when comparing bright and dim bursts, while the pre-peak slope hardly  
changes. Stronger bursts are thus more symmetric than weaker bursts
at a high confidence level. 
The very weakest bursts have a different shape (i.e., a different
stretched exponential index) of 
their average time profile as compared to brighter bursts.
This difference is too large to 
be explained by trigger effects and Poisson noise. We interpret these 
correlations as being the result of an intrinsic positive correlation 
between brightness and complexity of GRBs.
This  interpretation is directly confirmed by simulations as well as by
a morphological classification of bursts.
The fact that such a  correlation can be observed should impose new 
constraints on the distribution of GRBs  over luminosity distance.

\end{abstract}


\section{Introduction}

Although gamma-ray bursts (GRBs) are quite diverse in
their properties (e.g., Fishman \& Meegan 1995), clues to their origin
may be hidden in the average statistical properties of GRBs.
Mitrofanov et al. (1993, 1995) introduced a useful average function 
of light curves of GRBs. This is the average peak-aligned profile,
i.e., the sum of individual time profiles aligned at their highest peaks and 
normalized to their peak count rates. This function was used in searches for 
the time dilation effect (also see Norris et al.1994).
Stern (1996) found that the average post-peak time profile 
has a simple ``stretched'' exponential shape,
$F(t) \propto \exp[-(t/\td)^{1/3}]$, where $t$ is the
time measured from the peak of the
event, and $\td$ is a decay time constant. 
This fact, besides providing a possible clue for understanding the time 
variability 
of GRBs (Stern \& Svensson 1996), gives an excellent opportunity to study 
such effects as time dilation and time asymmetry of GRBs. These effects were 
previously studied using different methods 
(e.g.,  Nemiroff et al. 1994; Norris 1996 and references therein).

Stern (1996) used the stretched exponentiality of the average time profile
to study the time dilation of its post-peak slope.
Here we extend that study with a 
more accurate
procedure for the data analysis, using a larger sample of GRBs, and considering 
both the pre-peak and the post-peak slopes.
 Another recent advantage is the access to the
pulse avalanche model developed by Stern \& Svensson (1996), which successfully 
describes many 
statistical properties of GRBs including the stretched exponential shape 
of the average time profile, and, of particular importance for the present work,
the root-mean-square (rms) variance of individual time profiles. 
This means that we can rely on
this model when estimating the errors of stretched exponential fits, which  in 
turn gives us reliable estimates of the significance 
levels of any observed effects.

\section{Data Analysis} 

This work is based on the  publically available data in the Compton Gamma-Ray 
Observatory data archive in Goddard Space Flight Center. 
Our sample includes bursts
up to trigger number 3745 and contains 912 useful events. 
All time profiles are constructed using  64 ms time resolution data from the 
Large Area Detectors (LAD)  
together with 1.024 s resolution extensions to earlier and later times. 

We use the count rates summed over 
the four LAD energy channels,
covering the 25 -- 1000 keV energy range, as well as the count
rates in channels 2+3 (50 -- 300 keV). 
The advantage of the wider energy band is that it should reduce possible
effects of the spectral redshift if the photon spectrum has a "convex"
shape (which is usually the case).
On the other hand, channels 2 and 3 have a better signal to noise ratio. 
 We study the general behavior of
the average time profile using the wider energy band, and 
investigate the average shape of weak bursts using only channels 2 and 3.

The procedure of background fitting included a visual examination of all
bursts together with an analysis of all ambiguous signal variations 
to establish whether they came
from the same direction as the main burst. We discarded those bursts, for which
 a reliable
reconstruction of the signal for  50 s before and 125 s after the main peak
was not possible due to interfering background variations or data gaps. 
When possible we used two widely separated
background fitting windows ([$-120 \s$ , $-70 \s$] and
[$+200 \s$ , $+250 \s$] relative to the trigger).
In order to estimate the accuracy of the fitting procedure we measured 
residuals to the background fits for channels 2 and 3 in windows located 
approximately in the middle
of the fitted time interval avoiding possible contributions of a GRB signal.
We made this test for the 176 weakest events. 
The residuals normalized to the peak GRB fluxes are 
distributed in a symmetric Gaussian-like curve with an average value of
$+1.4 \times 10^{-3}$ and a rms deviation of 0.0187. The latter can be 
interpreted as the typical relative error of the fit for an individual profile. 
The statistical error of the
sum of relative residuals is $\pm 1.5 \times 10^{-3}$ - this is a rough estimate of the 
total uncertainty of the background for the average time profile of weak GRBs.  
The most difficult source of a possible bias for the weakest events are
background variations on the same time scale as GRBs which could be mixed
with the GRB signal. Our impression is that after we cleaned out events with
bad background such a bias is small. We cannot, however, quantify 
this conclusion. A test repeating the procedure using rescaled
events added to real samples of the background is necessary for such
a quantitative estimate.
For bright events the fitting dependent errors are negligible.

When sorting bursts into brightness groups, we used the peak fluxes 
for 64 ms time resolution in the BATSE-3 catalog (Meegan et al. 1996).
Peak alignment and peak normalization of bursts
were performed using count rates in 64 ms time resolution.
We developed
a flexible scheme where each count rate excess
over neighboring time intervals is tested for its statistical significance.
The shortest time scale in which the peak domination over its 
neighborhood is significant was used to localize the peak. 
The test for the brightness independency of the results obtained with such a
procedure is described below.

The parameters of an average time profile are derived from 
stretched exponential fits, 
$ F(t) = \beta \exp[-(t/t_0)^{\nu}]$, 
where, $\beta$ and $t_0$ are fitting parameters, and  the index $\nu$ was
free if its best fit value is of interest and was set to 1/3 when determining
the $t_0$. The value 1/3 does not necessarily have a special meaning, the 
$1 \sigma $ confidence interval is $\nu=0.34 \pm 0.02$. If we
set $\nu=0.32$ or $\nu=0.36$ then $t_0$ changes (by factor 1.5 for $\nu=0.36$),
however its ratio 
for different brightness groups remains the same within 4\% accuracy. 

Because the sample of all detected GRBs is too small to estimate the errors of
the average time profiles, we used the
{\it pulse avalanche} model of Stern \& Svensson (1996)
both to find a reasonable likelihood function for the fitting procedure and 
to estimate the statistical errors.
We found that the statistical errors are robust against variations 
of the model parameters. 
For the sample sizes used in our analysis, the statistical errors
 can be expressed as 
$\sigma(t_{\rm r,d})/t_{\rm r,d} = 0.201\sqrt{100/N}$,
$\sigma(\td + \tr)/(\td+\tr) = 0.196\sqrt{100/N}$, and
$\sigma(\td/\tr)/(\td/\tr) = 0.135 \sqrt{100/N}$,
where $\tr$ and $\td$ are the time constants  for the pre-peak (rising) 
and post-peak (decaying) slopes of the average time profile, respectively, 
and $N$ is number of events in a sample.
The better accuracy of $\td / \tr$ is caused by the
strong correlation between the two slopes.
The errors exceed the error estimates in 
earlier works (e.g., in Stern 1996, they were underestimated by a factor 1.5).
Large statistical errors when extracting time constants are natural 
if GRBs are produced by a near-critical process as is described 
by the pulse avalanche model of Stern \& Svensson (1996)
In the case of exact criticality, the time constants completely disappear. 
Note that our estimates of the statistical errors are, formally,
model-dependent. However, the model simultaneously describes
reasonably well the average time profile,
the autocorrelation function, and the rms variance of individual profiles 
(Stern \& Svensson 1996). Then it can be argued that our error estimates are 
also correct.

\section{The Average Time Profile for Different Brightness Groups} 

The average peak-aligned time profiles for three brightness groups 
are shown in Figure~1. 
Our stretched exponential fits are given in Table 1
and in Figure~2.
The rising (pre-peak) slope is steeper for all brightness groups,
but the asymmetry is increasing when going from the bright to the dimmest group.

In order to check a possible brightness dependent bias for $t_0$, we rescaled
strong bursts to  the
brightness ranges of the weaker samples, simulating noise and trigger operations. 
A parent sample consisting of the 268 brightest events has $\td = 0.503$.
After rescaling to the weaker samples 3, 4, 5, and 6, it has 
$\td =$ 0.500, 0.498, 0.510, and 0.525, respectively. Thus the possible bias is
less  than the statistical errors by an order of magnitude . The trigger
efficiencies for samples 3, 4, 5 and 6 are 0.993, 0.93, 0.77, and 0.38,
respectively.
 
Let us for the time being ignore  the weakest sample 6.
The strongest effect is the time dilation of $\td$
by a factor $1.85^{+0.82}_{-0.57}$ (90\% confidence interval)
if one compares samples 1 and 7.
The rejection level for a null hypothesis (i.e., no time dilation) 
is 0.998.
The rising slope, on the other hand, hardly changes and the 
variations of $\tr$ do not exceed the
statistical errors. This leads to a rising asymmetry for weaker samples
as quantified by the asymmetry ratio, $\td/\tr$. 
Comparing samples 1 and 7 gives
$(\td/\tr)_{\rm dim}/(\td/\tr)_{\rm bright} = 1.48^{+0.55}_{-0.32}$
($90\%$ confidence interval).
A careful estimate of the significance level using 
model simulations gives 0.985.
The time dilation of $\tr + \td$ being
$(\tr+\td)_{\rm dim}/(\tr+\td)_{\rm bright}$ =
$1.56^{+0.63}_{-0.36}$ (obtained by comparing samples 1 and 7)
has a statistical significance of 0.99. The value is in a good agreement
with recent results of Norris (1996), however the
definition of time dilation is too uncertain when strong shape-brightness 
correlations are observed.

Note that all effects are significant only relative to 
 the brightest end of the GRB brightness
distribution. If one removes the 64 brightest bursts, the effects
almost disappear - only marginally significant indications remain.
This is, however, natural as the brightest bursts cover a wide range 
of peak brightnesses on a logarithmic scale (see Fig.~2).
The asymmetry-brightness correlation cannot be explained as
a trigger effect, as the trigger efficiency is high throughout the brightness
range where we see this correlation.
It could not result from spectral redshift 
as our measurements of the time asymmetry in separate LAD energy channels
do not show any dependency on the energy channel (Stern et al. 1997).

The weakest sample 6 apparently has a different shape of its average 
time profile which is characterized by a stretched exponential index,
$\nu= 0.45$, while the other samples have $\nu \sim 0.33$ (in each energy 
channel). We tested whether 
the difference can results from trigger selection biases and
 from effects of Poisson noise.
As a reference weak sample, we used a sample (denoted as 8) including the 
197 weakest events
with the peak count rates in the range 0 -- 120 $\bin^{-1}$.
The best fit value of the 
stretched exponential index for sample 8 is $\nu = 0.47$ (channels 2 and 3).
The average trigger efficiency for strong events rescaled to the peak count rate
interval of sample 8 is 0.57. 
The index of the ``rescaled'' average time profile 
is $\nu = 0.37$. The deformations of the time profile after this procedure are 
of the  same sign as that distinguishing  samples 6 and 8 from brighter 
samples, but are much less than required to account for the difference in 
$\nu$.
 
To find the rejection level for the hypothesis that the peculiar $\nu$ = 0.47 for 
sample 8 is
the combined result of the trigger-associated bias and a statistical 
fluctuation,  
we performed model simulations taking into account Poisson noise and trigger
selection effects.
The resulting rejection level  
is close to 0.99 (3 out of 600 simulated samples gave  
$\nu > 0.47$). However we 
cannot rely on this estimate because we have other possible uncertainties.  

First, we could have both statistical and systematic errors in the background 
fit which can affect $\nu$. If we add to the average profile a constant
0.0015, which is our estimate of the fitting statistical error (see
the residuals  test in \S 2), then the rejection level drops to 0.97
(or, ifwe admit an error of 0.003, then $\nu = 0.447$ and the rejection
level becomes 0.94). 
Second, some profile-dependent biases could be accumulated in the various
steps of the ``data processing chain''. Such steps include: 
1) the classification of a trigger as a burst;
2) the inclusion of its 64 ms record into the database; 
and 3) the adding of the burst to the useful sample. 
According to Meegan (1996) the selection at steps 1) and 2) is not affected
by any features of the burst temporal structure. We admit that we may have
introduced a selection bias at step 3) at the level of 2 -- 4 events.
With these uncertainties we carefully characterize the significance of the
peculiar
$\nu$ for sample 8 as a ``$2 \sigma$ indication'' ($\sim$ 0.95). 
 
\section{Interpretation of Correlations} 

GRBs consist of a number of pulses of 
different durations but similar shapes. Let all these pulses have locally
independent sources of energy. Then if two pulses coincide in time, their 
amplitudes sum up. In a complex event hundreds of pulses are piling up, 
increasing the peak brightness by perhaps an order of magnitude. Then 
simple events are on average intrinsically weaker than complex events.
At the same time, simple events are asymmetric just because a single pulse 
in general is asymmetric with a sharp rise and a slower decay
(e.g., Norris et al. 1996). This asymmetry is washed
out in complex events where the position of the highest peak is more or less 
random among many overlapping pulses (a possible ``global'' 
asymmetry where brighter pulses tend to appear earlier remains). 
Furthermore, the sum of 
peak-aligned single pulse events gives $\nu \sim 0.5$.

To  demonstrate that this kind of correlation appears  in the 
pulse avalanche model, we 
simulated a large sample of ``bursts'' using a set of
model parameters fitted to the average time profile for the full
sample of all  real bursts. The results are given in Table~\ref{tab2},
where one can  see the tendencies of a larger asymmetry ratio,  $\td/\tr$, 
and a larger stretched exponential
index, $\nu$, for weaker events, as well as the effect of 
``time shrinking'', which is opposite to the observed time dilation. 
The latter implies a new positive correction for the observed time dilation.

\section{Complexity vs Observed Brightness} 

If correlations between the average time profile and brightness can be 
explained by complex bursts being intrinsically brighter than simple bursts,
then it is of interest to classify bursts according to their complexity
and then check whether any correlations exist between complexity and observed
brightness.
Lestrade (1994) using the number of runs "up" and "down" in GRB time profiles
as a measure of complexity did not find such correlations. Actually this is
a difficult problem: all events should be rescaled to the same low brightness
to avoid brightness-dependent biases, but then the effect is difficult to extract
due to the large Poisson noise. We found that the only way to detect
correlations is to concentrate on the simplest events dominated by a single
smooth pulse having a sharper rise and a slower decay. Such pulses
can be be recognized due to their "canonical" shape even at a high Poisson noise.

To make a brightness
independent classification we rescaled all 912 events to the same peak count 
rate range (125 -- 150 $\bin^{-1}$ in all energy channels), 
adding proper  Poisson noise. As it is important to conduct 
such a classification as a ``blind test'', we arranged the procedure in such a 
way that none of the test persons knew the actual peak amplitude of a 
displayed burst. All persons stated their decisions independently.

As a fine classification at high Poisson noise is impossible, we restricted 
the classification to two main classes: (1) ``simple'' single pulse events 
(see definition above), and (2) ``complex'' events. Doubtful and too short events
were not included into these classes.
    
The test was performed by three persons: A, B and C.
All test persons  demonstrated the same tendency: events classified as 
``simple'' are systematically dimmer than those classified  as ``complex''. 
A Kolmogorov-Smirnov test for the two  brightness distributions of
``simple'' and ``complex'' bursts gave the following consistency 
levels (i.e., the probability that the two distributions are drawn
from the same parent distribution): 
A -- $0.75\cdot 10^{-2}$, B -- 0.28, C -- $0.43\cdot 10^{-2}$, 
A $\times$ C (i.e., only those events
where A and C agreed) -- $0.38\cdot 10^{-2}$. 
The test directly confirms that complex events dominate the 
bright range and simple bursts dominate the weakest range (see Fig.~\ref{fig3}).

Some systematic bias of this test is still possible for the weakest bursts
having peak count rates below 125 $\bin^{-1}$
 as they have worse signal to noise ratios than rescaled events. 
The total number of such events is, however, small (74 out of 912 events).

We also performed a similar
computer test using a $\chi^2$ threshold as the criterium for the event to be
dominated by a single pulse. The parametrization of
Norris et al. (1996) was used as the fitting hypothesis for single "canonical" pulses
with the additional imposed requirement that the best fit
exponential decay time constant for an individual pulse is larger than the
corresponding rise time constant.  The result of
this test is surprisingly close to that of the visual test: "complex" events
are systematically brighter at the 0.997 confidence level. Details will be
given in Stern, Poutanen \& Svensson (1997).

\section{Conclusions and Discussion} 

The effects we confidently detect can be summarized as follows:
1.) The strongest bursts have peculiarly short time constants (confidence level 
$> 99\%$). 
2.) The strongest bursts have peculiarly small asymmetry
between the rising and the decaying slopes of the average time profile 
(confidence level
98 -- 99\%). 
3.) The weakest bursts ($\Fp \simless 0.8 \ph\cminvsq\secinv$) have a 
peculiar shape of
the decay slope of their average time profile, which can be characterized by a
large stretched exponential index (a formal estimate gives a 99\% confidence
level, a more realistic estimate is $2 \sigma$).
4.) Events consisting of a single pulse are systematically weaker than complex 
events (99.5\% confidence level).

Effect 1) can be interpreted as the widely discussed time dilation effect. 
Unfortunately other observed effects complicate its treatment. 
Therefore we can only state that after applying all possible 
corrections the value of the time dilation
lies between 1.5 and 4 and that an unknown fraction of this is due to
an intrinsic anti-correlation between brightness and duration (see Brainerd
1994).

Effects 2) and 3) have a common qualitative interpretation as being produced 
by an intrinsic
correlation between the number of pulses in a burst and its peak brightness. 
This 
interpretation is directly supported by 4) and by model simulations.

All these effects by themselves are not as important as the constraints
one can put on the distribution of GRBs over luminosity distance due to the fact
that we do observe these effects.
If some of GRB's properties correlate with their intrinsic luminosity, this
condition alone does not mean that we are able to observe such a correlation. 
The second necessary condition is that the distribution of GRBs over luminosity
distance is not a power law. This is necessary in order for the ratio of
intrinsically weak and intrinsically strong events to change at some apparent
brightness range.
As for the brightest range of GRBs this is more or less evident -- in the 
cosmological
scenario a transition to Euclidean scaling with a power law index $-3/2$ should
exist and this is indicated by the data (e.g., Meegan et al. 1996). 

The domination of intrinsically weak bursts in the weakest range would imply 
a less trivial turnover of the distance distribution towards larger distances.
Indeed, a relative deficit of intrinsically strong bursts at the weakest end 
of the brightness distribution indicates that we are sensing the edge 
of the spatial distribution of GRBs.
In a cosmological scenario this could imply  
strong evolution.
Effect 3) which leads to such a conclusion is measured near the threshold
of the detector sensitivity where one has to deal with many
possible biases. This makes effect 3) less significant than the other effects 
listed above. 

Nevertheless we believe that effect 3) is real as it is 
supported by effect 4). It certainly should be studied more attentively,
maybe using untriggered bursts in 1024 ms records.
As is often the case, the most difficult fraction
of the data can be the most valuable. 
We suggest that weak GRBs deserve more attention. 

\acknowledgments

We thank Cecilia Albertsson and Felix Ryde for helpful assistance and
Ed Fenimore and Charles Meegan for useful discussions.
We are grateful to the referee, Robert Nemiroff, for valuable remarks.
This research used data obtained from  
the HEASARC Online Service provided by the NASA/GSFC.
This study was supported by grants
from the Swedish Royal Academy of Sciences, the Swedish Natural Science 
Research Council, and a NORDITA Nordic Project grant.

\clearpage

\begin{table*}
\begin{center}
\begin{tabular}{l l l l l l l l}
\tableline
\# & Peak flux &  $N$ &  $\td$ & $\tr$  & $\td$ + $\tr$ &  $\td/\tr$ \\ 
\tableline
1 & 12.5 -- 200  & 64  & 0.42$\pm 0.10$ & 0.36$\pm0.09$ & 0.78 $\pm 0.18$ & 1.17 $\pm0.19$ \\
2 & 3 -- 12.5 & 193 & 0.60$\pm 0.08$ & 0.40$\pm0.06$ & 1.00 $\pm 0.13$ & 1.50 $\pm0.11$ \\
3 & 1.75 -- 3 & 159 & 0.75$\pm 0.11$ & 0.37$\pm0.06$ & 1.12 $\pm 0.16$ & 2.02 $\pm0.21$ \\
4 & 1 -- 1.75 &  241 & 0.72$\pm 0.09$ & 0.46$\pm0.06$ & 1.18 $\pm 0.14$ & 1.56 $\pm0.13$ \\
5 & 0.7 -- 1 & 139 & 0.85$\pm 0.14$ & 0.48$\pm0.08$ & 1.33 $\pm 0.21$ & 1.89 $\pm0.18$\\
6 & 0 -- 0.7  & 116 & 0.80$\pm 0.14$ & 0.37$\pm0.07$ & 1.17 $\pm 0.20$ & 2.16 $\pm0.26$ \\ 
7 & 0.75 -- 2.5   & 463 & 0.78$\pm 0.07$ & 0.45$\pm0.06$ & 1.22 $\pm 0.11$ & 1.73 $\pm0.10$\\
\tableline
\end{tabular}
\end{center}
\caption{Time constants of the average time profiles
}
\tablecomments{
Time constants, $\tr$ and $\td$ [$\s$],
of the stretched exponential fit to the
pre-peak (rising) and post-peak (decaying) average time profiles, respectively.
Peak fluxes
[$\ph\cminvsq\secinv$] taken from the BATSE data base for the sum of
channels 2 and 3.
$N$ is the number of bursts in a given brightness interval.
Errors obtained using the estimates in \S 2 correspond to $1\sigma$. 
}
\label{tab1}
\end{table*}

\begin{table}
\begin{center}
\begin{tabular}{l l l l l l l}
\tableline
Peak flux       & $\td$& $\tr$&  $\td$ + $\tr$ &  $\td/\tr$ & $\nu$\\ 
\tableline
0 -- $\infty$   & 0.66 & 0.41 &  1.07 & 1.61 & 0.34 \\
3 -- $\infty$   & 0.75 & 0.68 &  1.43 & 1.10 & 0.26\\
0.8 -- 3        & 0.72 & 0.50 &  1.22 & 1.44 & 0.31 \\
0 -- 0.8        & 0.50 & 0.19 &  0.69 & 2.63 & 0.48\\
\tableline
\end{tabular}
\end{center}
\caption{Fitting parameters of the average time profiles for 
simulated bursts}
\tablecomments{The fitting parameters, $\tr$, $\td$, and $\nu$, of the 
average time
profiles for simulated bursts in
different ``intrinsic brightness'' intervals. The amplitude of a
single pulse is sampled
uniformly in the $[0,1]$ interval. The stretched exponential index, $\nu$,
is given for the $0.75 < t^{1/3} <4$ fitting interval.  The parameter values 
are model-dependent, and are given only as an example.
}
\label{tab2}
\end{table}

\clearpage


\clearpage

\begin{figure}
\plotone{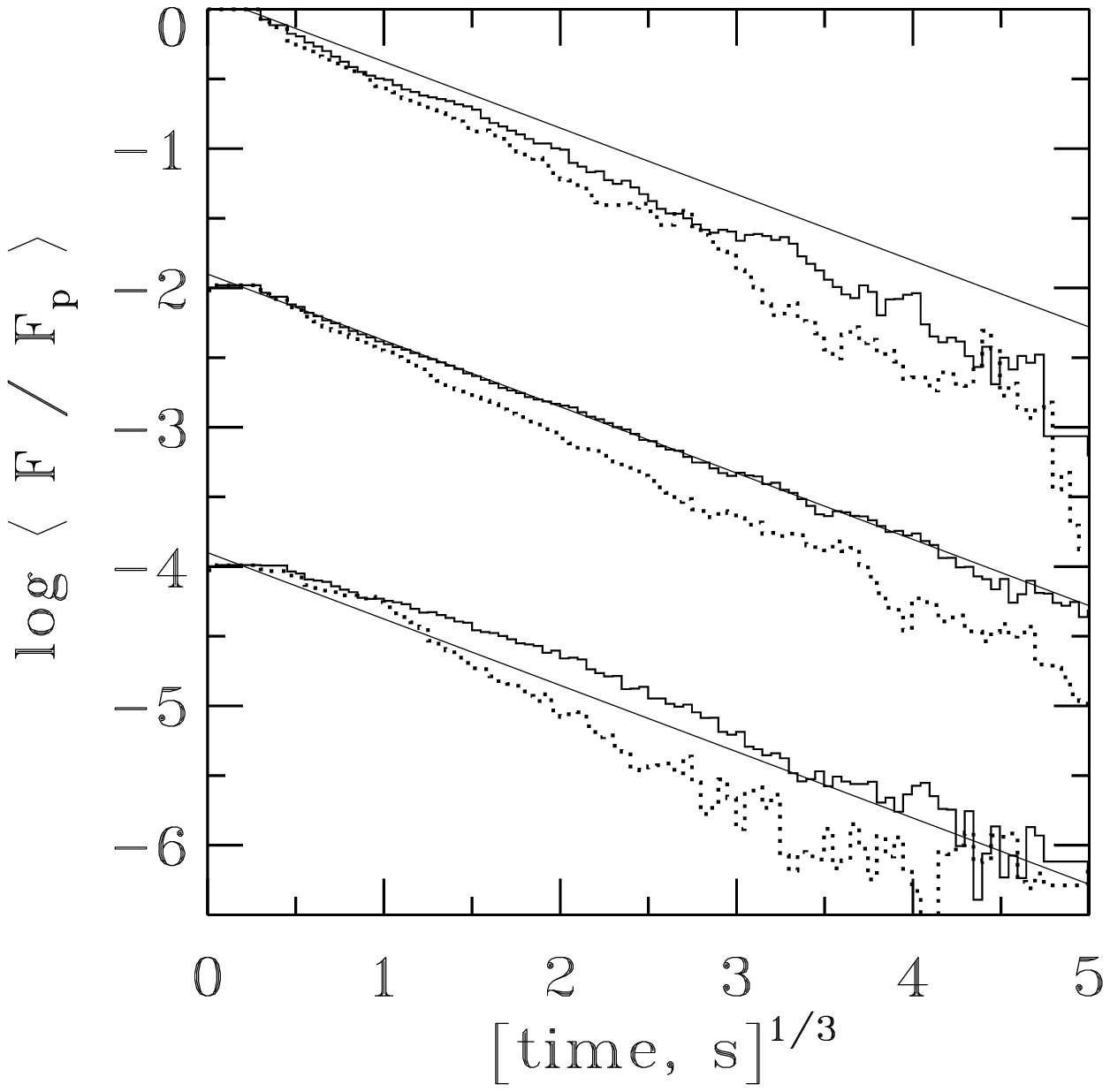}
\caption{Average peak-aligned time profiles for three brightness groups:
1) Peak photon flux $\Fp > 5 \ph\cminvsq\secinv$, 157 GRBs (upper curves);
2) $0.7 < \Fp < 5$, 630 GRBs (middle curves); and
3)  $\Fp < 0.7$, 116 GRBs (lower curves).
Average time profiles for medium and low brightness groups
are shifted downwards for clarity.
Solid and dotted curves represent the average post-peak
and pre-peak time profiles, respectively.
Straight lines show the best linear fit to the post-peak history
of the medium brightness group. The best fit time constants are 
$\tr$ = 0.35 s, $\td$ = 0.50 s  for the brightest group;
$\tr$ = 0.43 s, $\td$ = 0.74 s for the medium group;
and $\tr$ = 0.37 s, $\td$ = 0.80 s for the weakest group. See caption for 
Fig.~2 regarding the weakest sample.
}
\label{fig1}
\end{figure}

\begin{figure}
\plotone{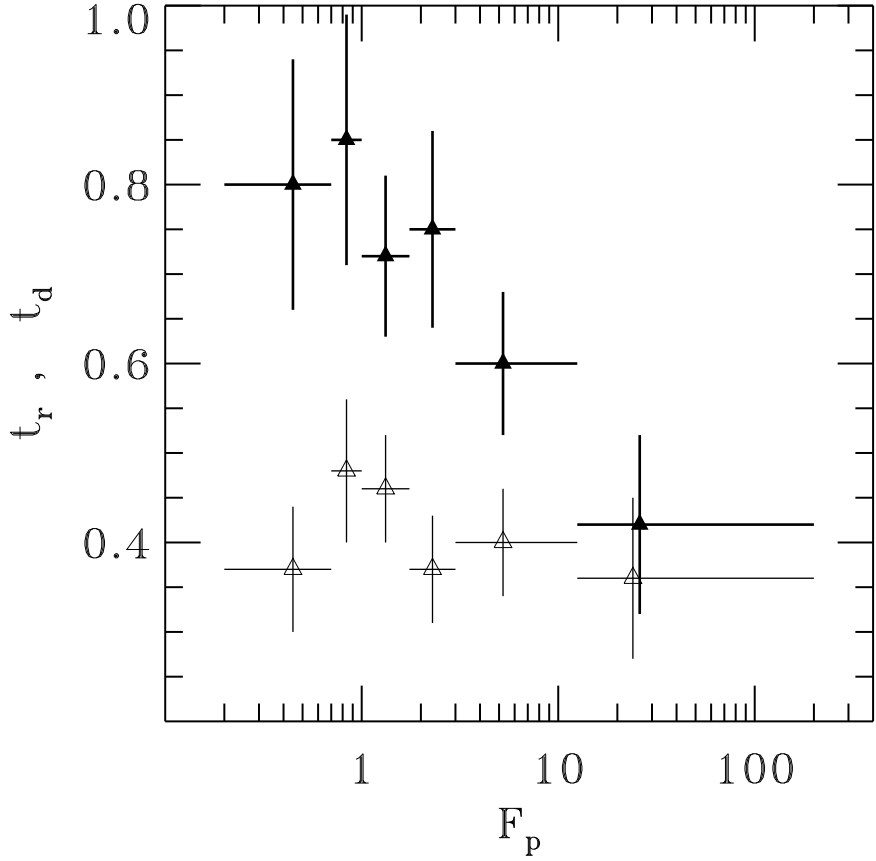}
\caption{Time constants, $\tr$ and $\td$,  vs. peak photon flux, $\Fp$, 
in $64 \ms$ time resolution.
Lower and upper crosses represent $\tr$ and $\td$ 
for the pre-peak (rising) and post-peak (decaying)  average
time profiles, respectively. Error bars of the time constants correspond
to $1\sigma$. Error bars in photon flux represent
the width of the brightness groups.
The values for the weakest sample (\# 6 in Table 1) are not trustworthy
because the average time profile apparently has a different shape as compared
to the stretched exponential function (see Fig. 1).
}
\label{fig2}
\end{figure}

\begin{figure}
\plotone{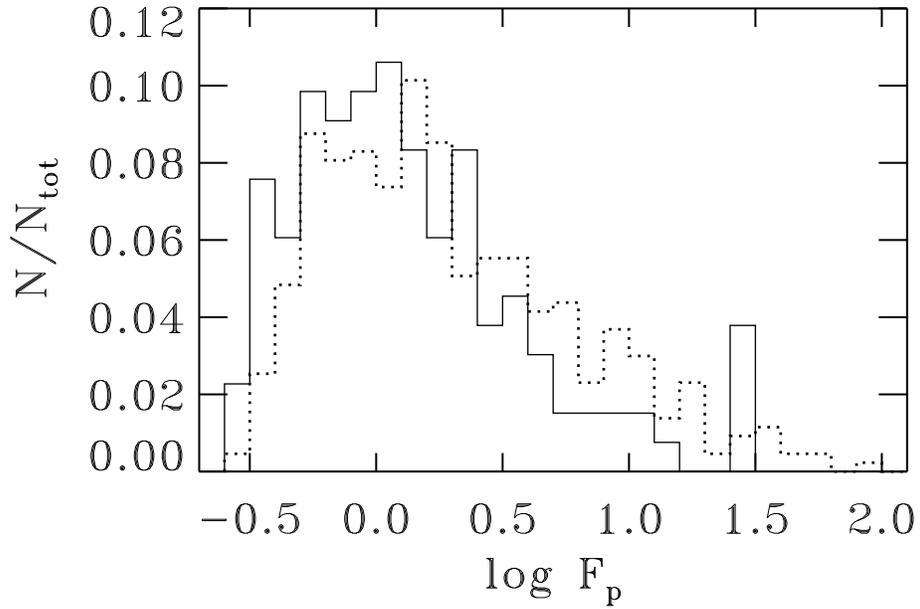} 
\caption{
The normalized peak brightness distributions of bursts classified
by both of the test persons A and C as ``one pulse events'' (solid histogram)
and as ``complex events'' (dotted histogram). $N_{\rm tot}$  is
the total number of bursts in respective class.
}
\label{fig3}
\end{figure}

\end{document}